\documentclass[baaa]{baaa}

\usepackage[pdftex]{hyperref}
\usepackage{subfigure}
\usepackage{natbib}
\usepackage{helvet,soul}
\usepackage[font=small]{caption}

\begin{document}


\journalvol{60}
\journalyear{2018}
\journaleditors{P. Benaglia, A.C. Rovero, R. Gamen \& M. Lares}


\contriblanguage{1}


\contribtype{2}

\thematicarea{1}

\title{Evolution of the stellar metallicities of galaxies}
\subtitle{in the EAGLE simulations}


\titlerunning{Stellar metallicities in the EAGLE simulations}


\author{M.E. De Rossi\inst{1,2}, R.G. Bower\inst{3}, A.S. Font\inst{4},
J. Schaye\inst{5}, T. Theuns\inst{3}}
\authorrunning{De Rossi et al.}


\contact{mariaemilia.dr@gmail.com}

\institute{
Universidad de Buenos Aires, Facultad de Ciencias Exactas y Naturales y Ciclo B\'asico Com\'un. Buenos Aires, Argentina
\and CONICET-Universidad de Buenos Aires, Instituto de Astronom\'{\i}a y F\'{\i}sica del Espacio (IAFE). Buenos Aires, Argentina
\and Institute for Computational Cosmology, University of Durham, Durham, Reino Unido
\and Astrophysics Research Institute, Liverpool John Moores University, Liverpool, Reino Unido
\and Leiden Observatory, Leiden University, Leiden, the Netherlands
}


\resumen{
Estudiamos la correlación entre masa estelar y metalicidad estelar en el conjunto
de simulaciones cosmológicas hidrodinámicas {\em Evolution and Assembly of GaLaxies and 
their Environments} (EAGLE).  A una dada masa estelar, las galaxias simuladas con 
metalicidades estelares menores muestran, en promedio, fracciones de gas mayores, tasas
de formación estelar específicas mayores y poblaciones estelares más jóvenes.
La retroalimentación al medio por núcleos activos de galaxias parece jugar
un rol importante en la determinación de la metalicidad estelar a masas altas.
En general, los sistemas simulados siguen una anticorrelación bien definida
entre metalicidad estelar y fracción de gas, la cual no evoluciona significativamente
con el corrimiento al rojo.  Todas estas tendencias son consistentes con hallazgos
previos respecto de la metalicidad del gas en regiones de formación estelar en EAGLE.}

\abstract{
We study the correlation between stellar mass and stellar metallicity in 
the Evolution and Assembly of GaLaxies and their Environments (EAGLE)
suite of cosmological hydrodynamical simulations.  At a given stellar mass,
simulated galaxies with lower stellar metallicities show, on average, higher gas fractions, higher
specific star formation rates and younger stellar populations.
Active galactic nuclei feedback seems to play an important role on the determination
of the stellar metallicity at high stellar masses.  
In general, simulated systems follow a well-defined anticorrelation 
between stellar metallicity and gas fraction, which does not evolve significantly with redshift. 
All these trends are consistent with previous findings regarding the metallicity of the star-forming gas
in EAGLE.
}



\maketitle

\section{Introduction}
\label{S_intro}

The study of the chemical abundances of galaxies is a topic of great
interest in the community as it can help to constrain galaxy formation
models \citep[e.g.][]{finlator2016}.
In the local Universe, there is a well-defined correlation between
gas-phase oxygen abundances (${\rm O/H}|_{\rm gas}$) and stellar masses ($M_*$) of galaxies in such a way 
that more massive systems are more metal-enriched \citep[e.g.][]{tremonti2004}.
The $M_* - {\rm O/H}|_{\rm gas}$ relation seems to evolve with redshift ($z$) in the sense
that galaxies of similar masses were less metal-enriched in the past \citep[e.g.][]{maiolino2008}.

In last years, different authors suggested that the $M_* - {\rm O/H}|_{\rm gas}$ relation
may be the projection on to two dimensions of a more fundamental relation (FMZR) between
$M_*$, ${\rm O/H}|_{\rm gas}$ and star formation rate (SFR) \citep[e.g.][]{mannucci2009}.
Other authors claimed that the FMZR might be a consequence of a more fundamental correlation
between $M_*$, ${\rm O/H}|_{\rm gas}$ and gas fraction ($f_{\rm gas}$) \citep[e.g.][]{bothwell2013}.

During the last decades, different theoretical models and simulations have tried 
to provide light into the origin of metallicity scaling relations 
\citep[e.g.][]{derossi2015, bahe2016}.
In a recent work, \citep{derossi2017} investigated different correlations between metallicities 
and other global properties of galaxies in the EAGLE simulations \citep{schaye2015}.
These authors focused mainly on the analysis of the star-forming (SF) gas metallicities, finding good
agreement between some observed trends and predictions from a high-resolution version of the simulations.
In the current article, we extend this previous work by analysing in more detail the
metallicities associated to the stellar component ($Z_*$) of EAGLE galaxies.

\begin{figure*}[!t]
  \centering
  \includegraphics[width=0.9\textwidth]{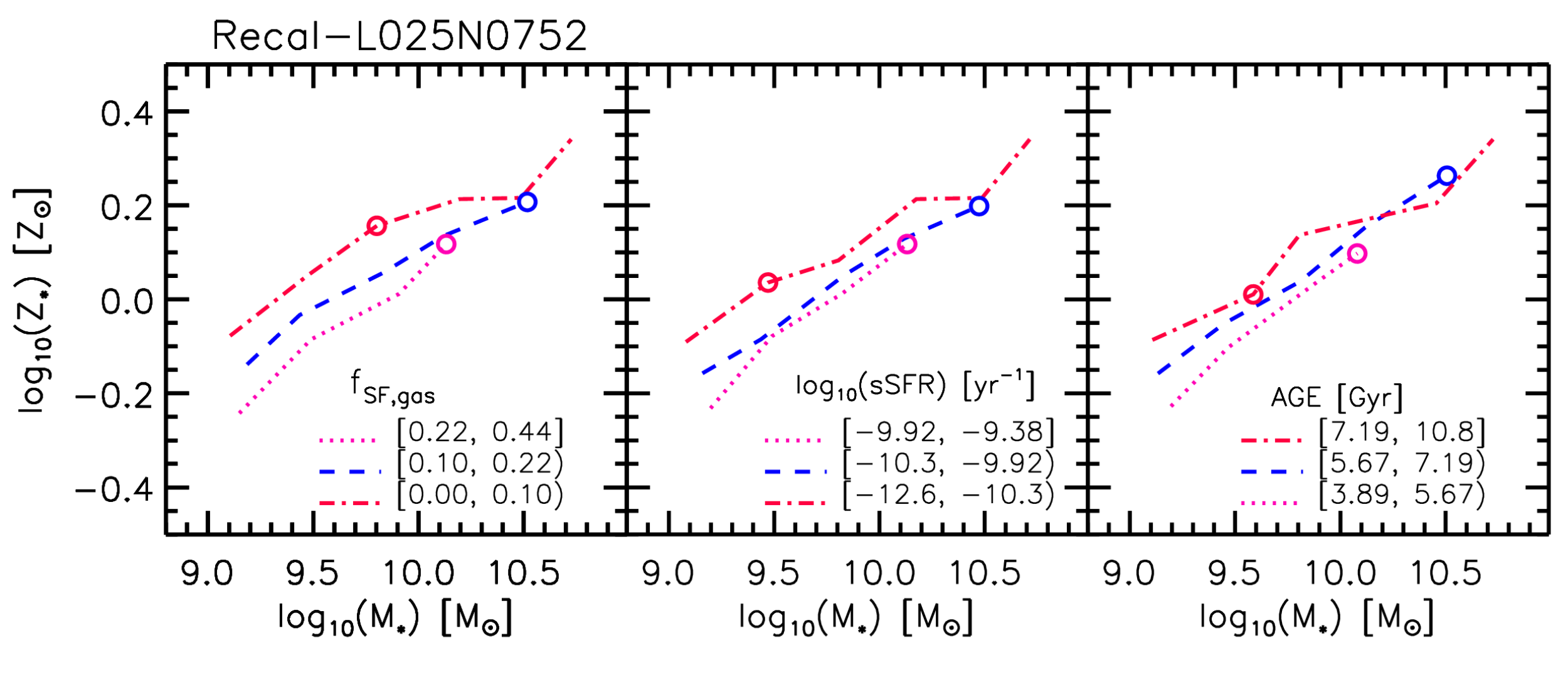}
  \caption{
Median $M_* - Z_*$ relation at $z=0$ binned in $f_{\rm SF,gas}$ (left panel),
sSFR (middle panel) and stellar mass-weighted mean age (right panel), as indicated in the figure.
All considered mass bins contain $N_{\rm bin} \ge 5$ galaxies; less populated bins ($ 5 \le N_{\rm bin} < 10$)
are marked with a circle.
}
  \label{scatter_MZR}
\end{figure*}

\section{Simulations}
\label{simus}

The EAGLE suite \citep{schaye2015, crain2015} is a set of cosmological hydrodynamical simulations run in
cubic, periodic volumes ranging from 25 to 100 comoving Mpc. 
These simulations were run using a modified version of the {\sc GADGET-3} code and
adopting a $\Lambda$-CDM flat cosmology: ${\Omega}_{\Lambda}=0.693$, 
${\Omega}_{\rm m}=0.307$, ${\Omega}_{\rm b}=0.04825$ and $h=0.6777$
\citep{planck2014}. The simulations implement state-of-the-art numerical
techniques and subgrid models for radiative cooling, star formation, stellar mass
loss and metal enrichment, energy feedback from star formation and active galactic
nuclei (AGN) feedback, among others.

Unless otherwise specified, in this work, we present results obtained from the high-resolution simulation 
Recal-L025N0752, which predicts metallicity scaling relations in better agreement
with some observed trends.
This simulation tracks the evolution of an initial number of $752^3$ particles
per species within a box of side-length of L=25 comoving Mpc and assumes
recalibrated parameter values to improve the match to the observed $z \sim 0$ galaxy stellar
mass function when increasing the resolution (see \citealt{schaye2015}, for details).





\section{Results}

\citet{derossi2017} found that EAGLE-Recal-L025N0752 simulation predicts a correlation
between $M_*$ and $Z_*$ consistent with the observed behaviour (their Fig. 5).
In this section, we explore secondary dependences of $Z_*$ at a given
$M_*$.  We also analyse the role of AGN feedback on $Z_*$ at high masses. 

\subsection{The $Z_* - M_*$ relation and its scatter}

In Fig. \ref{scatter_MZR}, we show the $M_* - Z_*$ relation binned
according to SF gas fraction ($f_{\rm SF,gas}$) (left panel), specific SFR (sSFR, middle panel) 
and stellar mass-weighted mean age (right panel), as indicated in the figure. 
The gas fraction is calculated as $M_{\rm SF,gas} / (M_{\rm SF,gas} + M_{*})$, where
$M_{\rm SF,gas}$ denotes the star-forming gas component.
At a given $M_*$, lower $Z_*$ can be associated, on average, to higher $f_{\rm SF,gas}$,
higher sSFR and younger stellar populations.  
Similar trends were obtained by \citet{derossi2017} for SF gas-phase oxygen abundances 
(${\rm O/H}|_{\rm SF, gas}$) at $M_* < 10^{10.3}~{\rm M}_{\sun}$ (see their Fig. 8).  
We note, however, that the secondary dependences obtained for $Z_*$ at a given mass 
are weaker than those previously found for ${\rm O/H}|_{\rm SF, gas}$ in a similar mass range.  
This behaviour is consistent with a scenario in which the infall of metal-poor gas in low-mass galaxies
leads to higher sSFRs and younger stellar populations in these systems.

\begin{figure}[!t]
  \centering
  \includegraphics[width=0.45\textwidth]{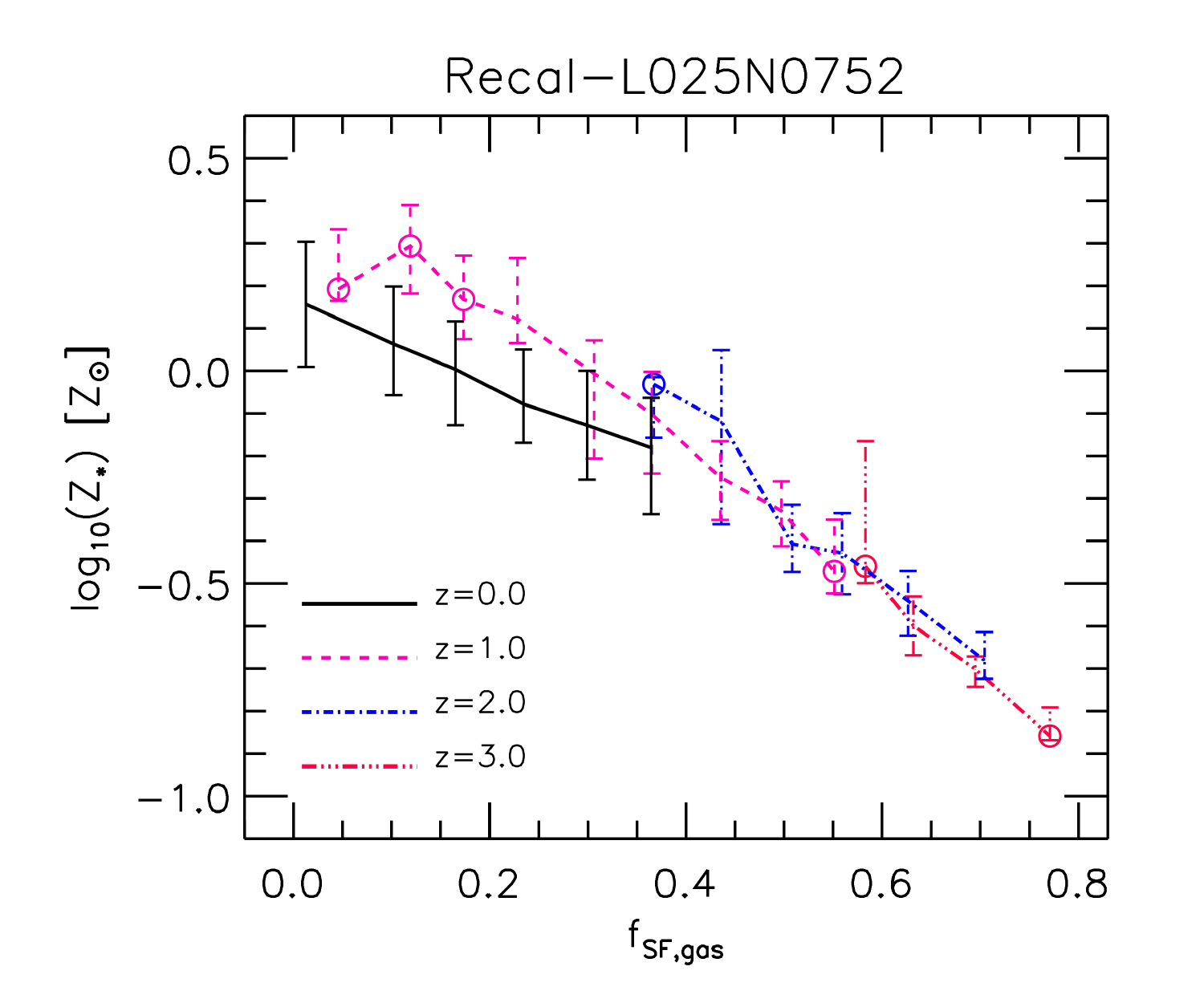}
  \caption{
Stellar metallicity as a function of star-forming gas fraction
for EAGLE galaxies at different $z$, as indicated in the figure.
Curves depict the median relation and error bars, the 25th and 75th percentiles.
The number of galaxies per bin is $N_{\rm bin} \ge 7$, with circles indicating less populated
bins ($N_{\rm bin} = 7-9$).
}
  \label{Zs_vs_fg}
\end{figure}

According to our findings, $Z_*$ exhibits a strong anticorrelation with $f_{\rm SF,gas}$ (Fig. \ref{Zs_vs_fg}).
The $f_{\rm SF,gas} - Z_*$ relation  does not evolves significantly with $z$
and shows a moderate scatter.
As $f_{\rm SF,gas}$ increases from $\approx 0$ to $\approx 0.8$, $Z_*$ decreases by more than 1~dex. 
These results are consistent with those obtained by \citet{derossi2017} in the case of the
SF-gas metallicities, which are also consistent with the so-called ``universal metallicity
relation'' found by \citet{zahid2014a} in the context of empirical-constrained analytical models.

\subsection{AGN feedback effects}

In order to explore the impact of AGN feedback on the $M_* - Z_*$ relation, we need
to focus on the trends at high masses and, thus, we employed the intermediate
resolution simulations L050N0752.  These simulations were run in a volume of side length of 50 comoving Mpc
including $752^3$ particles.  Four subsets of simulations were studied comprising
four models featuring variations of the temperature increment of stochastic AGN heating
(${\Delta}T_{\rm AGN}$): NOAGN (AGN effects suppressed entirely), 
Ref (reference model, ${\Delta}T_{\rm AGN}=10^{8.5}$~K), AGNdT8 (${\Delta}T_{\rm AGN}=10^{8}$~K)
and AGNdT9 (${\Delta}T_{\rm AGN}=10^{9}$~K) (see \citealt{schaye2015} and \citealt{crain2015}, for
a description of the simulations).

\begin{figure}[!t]
  \centering
  \includegraphics[width=0.45\textwidth]{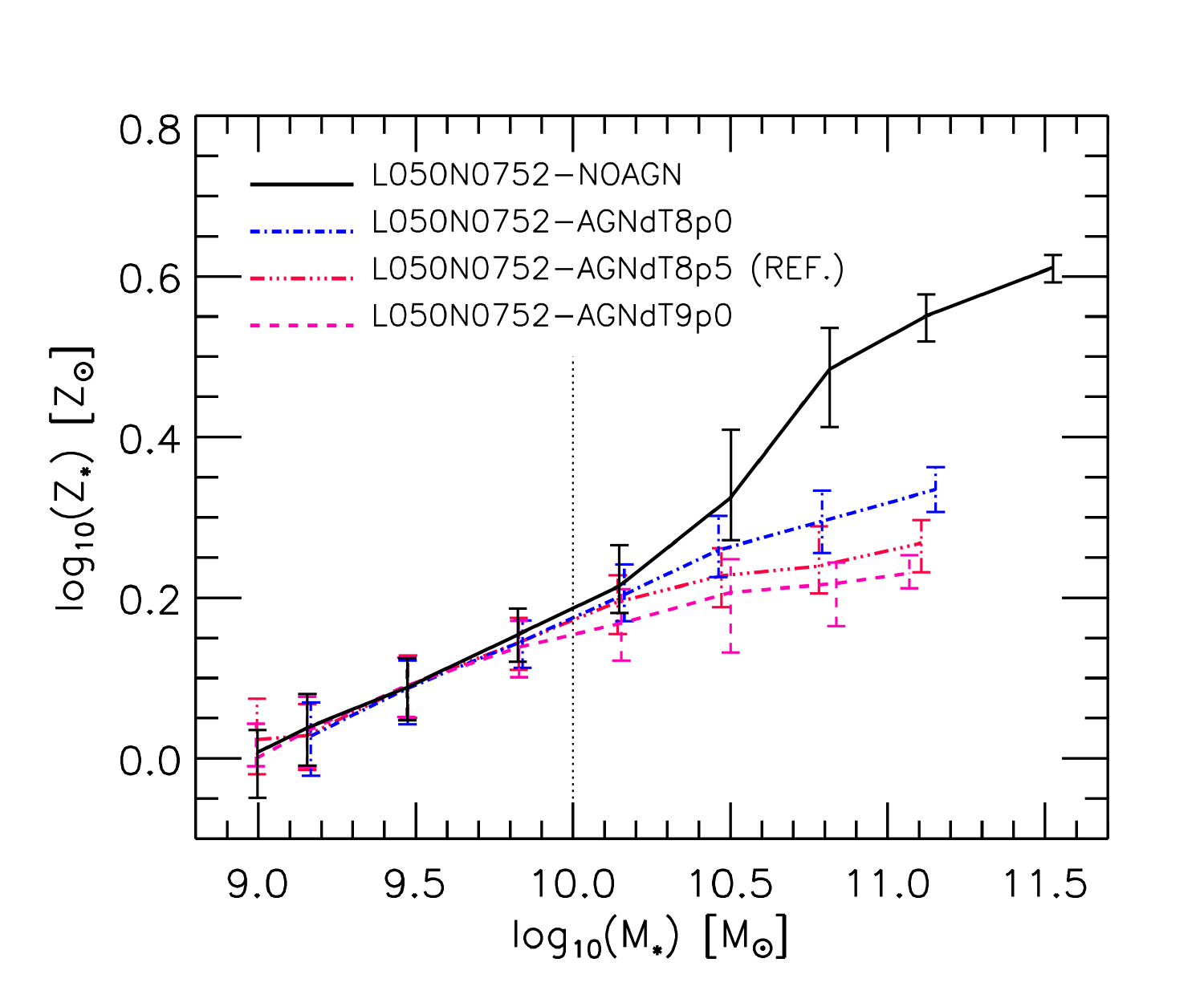}
  \caption{
Simulated $M_* - Z_*$ relations at $z=0$ for different models.
Results obtained from simulations
``L050N0752'' with different AGN feedback parameters are presented:
NOAGN (AGN feedback suppressed entirely),
AGNdT8 ($\Delta T_{\rm AGN} = 10^{8}$~K),
reference model ($\Delta T_{\rm AGN} = 10^{8.5}$~K) and
AGNdT9 ($\Delta T_{\rm AGN} = 10^{9}$~K).
Note that AGN effects set in above $M_* \sim 10^{10}~{\rm M}_{\sun}$ (dashed vertical line).
}
  \label{MZR_AGN}
\end{figure}

Fig. \ref{MZR_AGN} shows that AGN feedback plays an important role on the determination of 
stellar metallicities at high masses, at least in these simulations.
The slope of the $M_* - Z_*$ relation decreases
with ${\Delta}T_{\rm AGN}$ and, when AGN feedback is completely suppressed, $Z_*$ increases by up
to $0.3$ dex at $M_* \sim 10^{11}~{\rm M}_{\sun}$.  As discussed in detail in \citet{derossi2017},
AGN feedback leads to a decrease in the metal content of galaxies by quenching the star formation process
and generating the ejection of metal enriched gas from galaxies. Net metal dilution 
might also have a (minor) impact on the metal enrichment of massive AGN-host galaxies.


\section{Summary}

We studied stellar metallicity scaling relations in the EAGLE suite
of cosmological hydrodynamical simulations. We focused mainly on the high-resolution
simulation run that implements the so-called recalibrated model.
At a given $M_*$, simulated galaxies with lower $Z_*$ tend to
have higher SF gas fractions, higher sSFR and younger stellar populations.
These trends are stronger at $M_* < 10^{10.3}~{\rm M}_{\sun}$.
In general, we found a strong anticorrelation between $Z_*$ and SF gas fraction,
in agreement with the existence of the so-called ``universal metallicity relation'' reported by
\citet{zahid2014a}.

To explore the impact of AGN feedback, we employed intermediate resolution simulations
in which the AGN feedback temperature is varied.
Our findings suggest that AGN feedback plays an important role on the determination of
 the slope of the $M_* - Z_*$ relation at $M_* > 10^{10}~{\rm M}_{\sun}$.
Thus, the study of the mass-metallicity relation at high masses could help to constrain AGN feedback models.

For more details about this work and results for SF gas-phase metallicities, 
the reader is referred to \citet{derossi2017}.

%

\begin{acknowledgement}
We acknowledge support from PICT-2015-3125 of ANPCyT, PIP 112-201501-00447
of CONICET, UNLP G151 of UNLP (Argentina) and
STFC consolidated and rolling grants ST/L00075X/1 (Durham, UK).
This work was supported by the Netherlands Organisation for Scientific Research (NWO), through VICI grant
639.043.409, and the European Research Council under the European Union's Seventh Framework Programme
(FP7/20072013)/ERC Grant agreement 278594-GasAroundGalaxies.
We acknowledge support from the European Commission's Framework Programme 7,
through the Marie Curie International Research Staff Exchange Scheme LACEGAL (PIRSES-GA-2010-269264).
We acknowledge the Virgo Consortium for making their simulation data available.
The EAGLE simulations were performed using the DiRAC-2 facility at Durham, managed by the ICC,
and the PRACE facility Curie based in France at TGCC, CEA, Bruy\`res-le-Ch\^atel.
This work used the DiRAC Data Centric system at Durham University, operated by the Institute for Computational Cosmology on behalf of the STFC DiRAC HPC Facility (www.dirac.ac.uk). This equipment was funded by BIS National E-infrastructure capital grant ST/K00042X/1, STFC capital grants ST/H008519/1 and ST/K00087X/1, STFC DiRAC Operations grant ST/K003267/1 and Durham University. DiRAC is part of the National E-Infrastructure.
\end{acknowledgement}


\bibliographystyle{baaa}
\small
\bibliography{biblio}
 
\end{document}